\documentclass[pra,aps,twocolumn,amsmath,amssymb,floatfix,pra,reprint,footinbib,superscriptaddress,longbibliography]{revtex4-2}
\usepackage{graphics}      % standard graphics specifications
\usepackage{graphicx}      % alternative graphics specifications
\usepackage{longtable}     % helps with long table options
\usepackage{url}           % for on-line citations
\usepackage{bm}    
\usepackage{braket}        % special 'bold-math' package
\usepackage{xcolor}
\usepackage{float}  
\usepackage{natbib}
\usepackage{graphicx}
\usepackage{comment}
\usepackage{siunitx}
\usepackage{appendix}
\usepackage{amsmath,amssymb,epsfig,color}

\usepackage[most]{tcolorbox}

\newcommand{\nn}{\nonumber}

\newcommand{\be}{\begin{equation}}
\newcommand{\ee}{\end{equation}}
\newcommand{\bea}{\begin{eqnarray}}
\newcommand{\eea}{\end{eqnarray}}

\begin{document}

%\title { Chiral Emission of Giant Atom without Breaking Time-Reversal Symmetry}
\title {Chirality, Nonreciprocity and Symmetries for a Giant Atom}

\author{Luting Xu}
\affiliation{Center for Joint Quantum Studies and Department of Physics, School of Science, Tianjin University, Tianjin 300072, China }
%\affiliation{School of Science, Nanjing University of Science and Technology, Nanjing 210094, China}

\author{Lingzhen Guo}
\thanks{lingzhen\_guo@tju.edu.cn}
\affiliation{Center for Joint Quantum Studies and Department of Physics, School of Science, Tianjin University, Tianjin 300072, China }

%\author{Florian Marquardt}
%\affiliation{Max Planck Institute for the Science of Light, Staudtstrasse 2, 91058 Erlangen, Germany}
%\affiliation{Department of Physics , University of Erlangen-Nuremberg, Staudtstrasse 5, 91058 Erlangen, Germany}

\begin{abstract}
Chiral and nonreciprocal quantum devices are crucial for signal routing and processing in a quantum network. In this work, we study the chirality and nonreciprocity of a giant atom coupled to a one-dimensional waveguide. We clarify that the chiral emission of the giant atom is not directly related to the time-reversal symmetry breaking but to the mirror-symmetry breaking. We propose a passive scheme to realize the chiral emission of a giant atom without breaking time-reversal symmetry by extending the legs of the giant atom. We find the time-reversal symmetry breaking via nonuniform coupling phases is artificial and thus cannot result in nonreciprocal single-photon scattering for the giant atom. The nonreciprocity of the giant atom can be obtained by the external dissipation of the giant atom that truly breaks the time-reversal symmetry. Our work clarifies the roles of symmetries in the chirality and nonreciprocity of giant-atom systems and paves the way for the design of on-chip functional devices with superconducting giant atoms.
\end{abstract}

\date{\today}

\maketitle

\section{Introduction} 
Functional devices for chiral or nonreciprocal information processing in a quantum network usually need bulky and lossy magnetic materials with strong magnetic fields \cite{caloz2018prapp,abdo2019nc} that are off-chip disadvantageous for integration. For building a scalable quantum network, on-chip chiral or nonreciprocal interfaces to superconducting circuits with Josephson junctions (JJs) are also of immense technological interest. 
Most of the existing proposals for designing on-chip chiral or reciprocal devices break \textit{time-reversal} ($\cal {T}$) symmetry with tailored active control strategies \cite{hoi2011prl,sliwa2015prx,abdo2019nc} via dynamic modulation of dc superconducting microwave quantum interference devices \cite{karmal2011np,abdo2014prl,estep2014np,kerckhoff2015prapp,metelmann2018pra,roushan2017np,chapman2-19prapp}. %
On the other hand, passive functional devices that can simplify the experimental implementations are of great importance but remain elusive. A JJ ring threaded by a constant magnetic flux breaking $\cal {T}$-symmetry \cite{koch2010pra} is susceptible to charge noise and limited by bandwidth. By carefully exploiting engineered circuits with high impedance \cite{richman2021prx}, the limited bandwidth can be increased substantially, and an on-chip tunable broadband charge-insensitive chiral system has been developed \cite{zhang2021prl}.

Recently, giant atoms with superconducting circuits offer new possibilities to engineer on-chip scalable chiral and nonreciprocal functional devices \cite{li2023tunable,wang2021prl}. For example, a time modulation of the coupling between giant atom and waveguide \cite{zhang2021prl,Wang2022qst,wang2024prr,joshi2023prx} can result in chiral microwave photonics~\cite{Wang2022qst,wang2024prr,joshi2023prx,Guimond2020,Kannan2023nat,du2022prl,vega2023prr}, where the $\cal {T}$-symmetry breaking via the coupling phase imparted by the dynamical modulation is considered the key for chirality.
Meanwhile, a giant-atom system composed of two superconducting transmon qubits coupled to a common waveguide allows two delocalized orthogonal excitations to emit (and absorb) photons propagating in opposite directions ~\cite{Kannan2023nat}, where the $\cal {T}$-symmetry is \textit{not} broken because the unidirectional interaction is realized with two degenerate levels~\cite{Guimond2020,gheeraert2020pra}.  

\begin{figure}
  \centering
 \centerline{\includegraphics[width=0.9\linewidth]{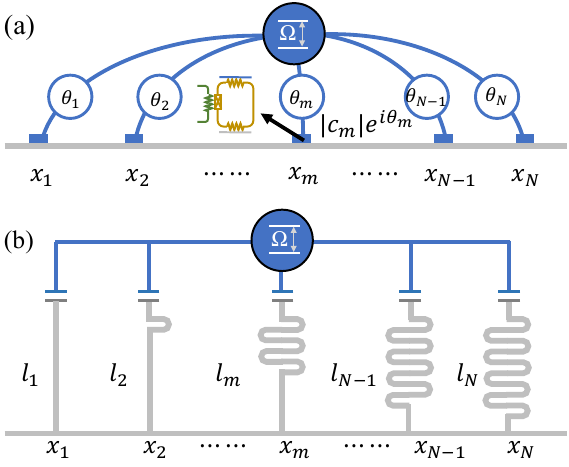}}
  \caption{{\bf Sketch of giant atom.}
  {\bf (a)} A two-level giant atom (blue) with transition frequency $\Omega$ coupled to a transmission line (gray) via multiple points $x_m$ with complex coupling strength $c_m=|c_m|e^{i\theta_m}$, where the coupling phases $\theta_m$ is imparted by a time-modulated active JJ loop (insert).  {\bf (b)} A passive giant-atom architecture with elongated lengths $l_m$ of coupling legs implemented via meandering transmission lines in superconducting circuits. The photon accumulates a propagating phase $\theta_m=\Omega l_m$ across each coupling line equivalent to the coupling phases shown in (a). 
   }\label{Fig-system}
\end{figure}

In this work, we find that the chirality of a giant atom is not directly related to the $\cal {T}$-symmetry breaking but to the mirror or parity ($\cal{P}$)  symmetry breaking. We propose two schemes, that break the $\cal {P}$-symmetry while preserving the $\cal {T}$-symmetry, to realize the chiral emission of a giant atom, which can be tuned on demand solely via the atomic frequency.
We further clarify that the $\cal {T}$-symmetry breaking in the giant-atom system via dynamic modulation of coupling is \textit{artificial} because it can be mapped to $\cal {T}$-symmetry preserved system. This can explain the previous confusion about why there is no nonreciprocity in such $\cal {T}$-symmetry broken system~\cite{Chen2022cp}. The nonreciprocity of the giant atom can be obtained by introducing external dissipation to the giant atom that truly breaks the $\cal {T}$-symmetry.

\section{Model Hamiltonian} 
We consider the model of a two-level giant atom coupled to the waveguides with multiple coupling points as illustrated in Fig.~\ref{Fig-system}(a). The giant atom interacts with the one-dimensional (1D) waveguide via $N$ coupling points with the total Hamiltonian 
\begin{eqnarray}\label{TotalH}
\hat{H}_s&=&\hbar\Omega\hat{\sigma}^+\hat{\sigma}^-
+\int_{-\infty}^{+\infty}\  dk\  \hbar\omega_k\
\hat{a}_k^\dag\hat{a}_{k}\\
&&+\sum_{m=1}^{N}\int_{-\infty}^{+\infty}
\Big(c_{m}e^{ikx_{m}}\hat{a}_k\hat{\sigma}^++{\rm h.c.}\Big)\sqrt{\omega_k}dk.\nonumber
\end{eqnarray}
%
\begin{comment}
\begin{eqnarray}\label{TotalH}
H_s&=&\hbar\Omega(\hat{b}^\dag\hat{b}+\frac{1}{2})-\kappa(\hat{b}+\hat{b}^\dag)^4
+\int_{-\infty}^{+\infty}\  dk\  \hbar\omega_k\
\hat{a}_k^\dag\hat{a}_{k}\nonumber\\
&&+\sum_{m=1}^{N}\int_{-\infty}^{+\infty}
\Big(c_{m}e^{ikx_{m}}\hat{a}_k\hat{b}^\dag+{\rm H.c.}\Big)\sqrt{\omega_k}dk.
\end{eqnarray}
\end{comment}
%
Here, we have defined the atomic operators $\sigma_+\equiv |e\rangle\langle g|$ and $\sigma_-\equiv |g\rangle\langle e|$, where $|e\rangle$ ($|g\rangle$) is the excited (ground) state of the atom, and  $\Omega$ is the atomic transition frequency. The field operators $\hat{a}_k$ represents the bosonic mode with wavenumber $k$ in the 1D waveguide satisfing $[\hat{a}_k,\hat{a}^\dagger_{k'}]=\delta(k-k')$. We assume the waveguide has a linear dispersion relation $\omega_k=|k|v$ with  $\omega_k$ the frequency and $v$ the propagating velocity. The interaction term in the second line of Eq.~(\ref{TotalH}) describes the Jaynes-Cummings (JC) coupling between the giant atom and the waveguide via multiple coupling points labeled by $x_m$ with $m=1,2\cdots, N$. The complex parameter $c_m=|c_m|e^{i\theta_m}$ represents the coupling strength at coupling position $x_m$, where the coupling phases $\theta_m$ can be imparted by a time-modulated JJ loop at each coupling point~\cite{zhang2021prl, Wang2022qst} as illustrated by the insert in Fig.~\ref{Fig-system}(a).

\section{Spontaneous emission} 
For the process of spontaneous emission, there is only one excitation either in the atomic state or in the waveguide as the JC interaction conserves the particle number.  We thus restrict the total state of the system to the single excitation subspace given by
\begin{eqnarray}\label{eq-TotalWave}
|\Psi(t)\rangle&=&\beta(t)|e,{\rm vac}\rangle+\int dk  \alpha_k(t) a^\dagger_k|g,{\rm vac}\rangle,
\end{eqnarray}
where $|\rm{vac}\rangle$ represents the vacuum state in the waveguide, and the integral term describes the state of a single boson propagating in the waveguide. Given the solution of the atomic probability amplitude $\beta(t)$, the dynamics of bosonic field function $\varphi(x,t)$ at position $x$ and time $t$ in the waveguide is  (see Ref.~\cite{guo2020prr} or Appendix \ref{sec-model})
\begin{eqnarray}\label{eq-varphi}
\varphi(x,t)&\equiv&\frac{1}{\sqrt{2\pi}}\int_{-\infty}^\infty dk
 e^{ikx}\ \alpha_k(t)\\
&=&-i\sqrt{\frac{\Gamma }{v}} \sum_{m}c_{m}^*\beta(t-\frac{|x-x_m|}{v})\Theta(t-\frac{|x-x_m|}{v}).\nonumber
\end{eqnarray}
Here, we have introduced the parameter $\Gamma\equiv\frac{2\pi\Omega}{\hbar^2 v}$, and
$\Theta(x)$ is the Heaviside step function.
 The field intensity in the waveguide $p(x,t)=| \varphi(x,t)|^2$ describes the probability density of a propagating photon.
The accumulated probability of the left/right propagating photon $\mathcal{I}_{L/R}(t)$ can be calculated as follows
\begin{eqnarray}\label{eq-IR}
\mathcal{I}_{L}(t)=v\int_0^{t}p(x_1,t') dt',\ \ \mathcal{I}_R=v\int_0^{t} p(x_N,t') dt'.  
\end{eqnarray}
We call the emission is \textit{chiral} if the accumulated probabilities to opposite directions are not identical, i.e., $\mathcal{I}_{L}(t)\neq\mathcal{I}_{R}(t)$,
and quantify the \textit{chirality} by
\begin{eqnarray}\label{eq-chirality}
\mathcal{C}(t)\equiv\frac{\mathcal{I}_L(t)-\mathcal{I}_R(t)}{\mathcal{I}_L(t)+\mathcal{I}_R(t)}.
\end{eqnarray}
%%%change the label
Note that the chirality defined here is time-dependent.

\begin{figure}
  \centering
 \centerline{\includegraphics[width=0.9\linewidth]{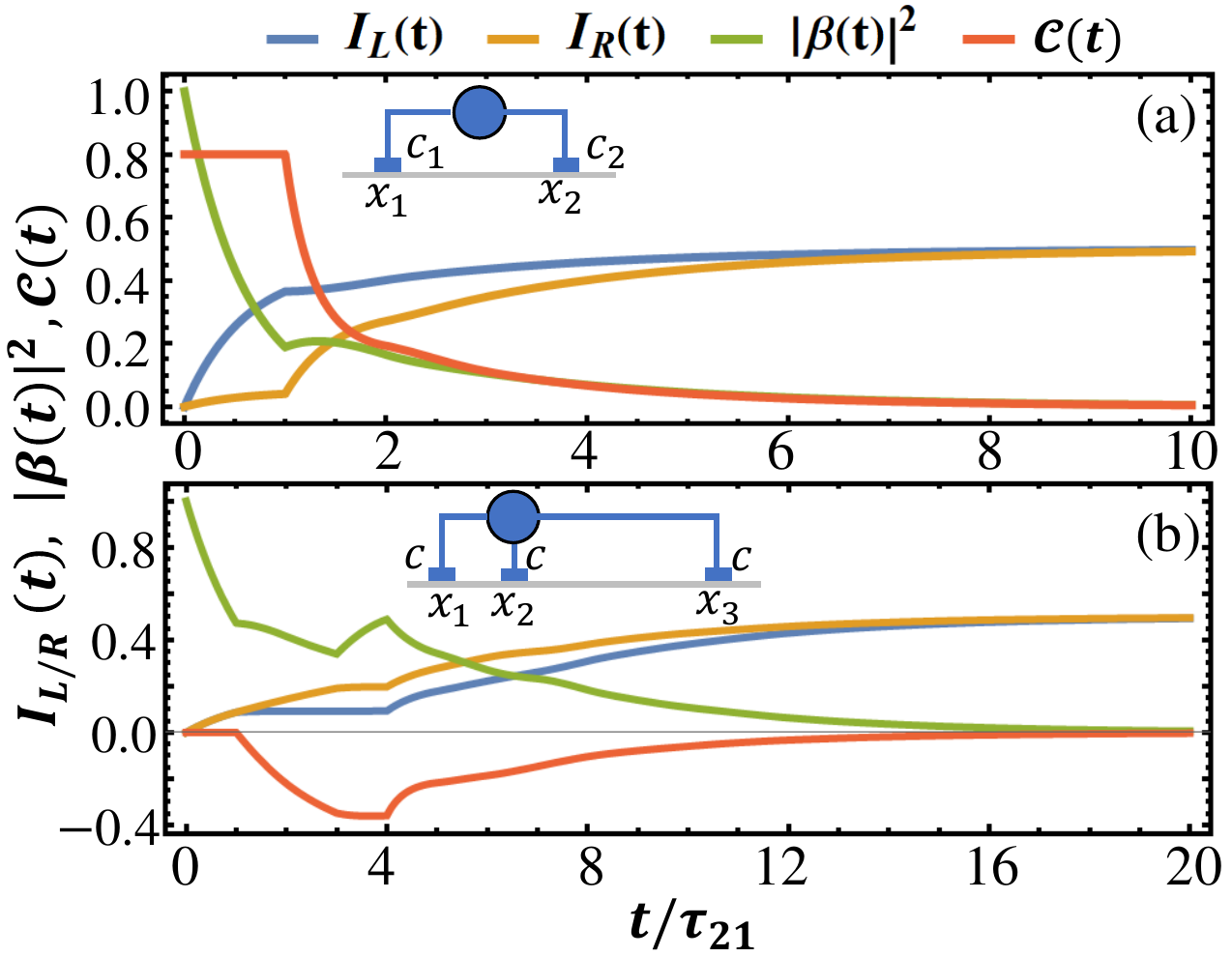}}
  \caption{{\bf Chirality of a non-Markovian giant atom.}
  {\bf (a)}  Spontaneous emission of a two-leg giant atom with real-valued but different coupling parameters $c_1=3c_2$ in the non-Markovian regime $\gamma \tau_{21}=0.5$ with decay rate $\gamma\equiv2\Gamma c_1c_2$ and propagating phase $\Omega\tau_{21}=\pi$ with $\tau_{21}\equiv (x_2-x_1)/v$ the delay time between two legs. Here, $I_{L/R}(t)$ are the accumulated probabilities of the emitted photon to the left/right directions, $\mathcal{C}(t)$ is the chirality defined by Eq.~(\ref{eq-chirality}), and $|\beta(t)|^2$ is the population of the excited giant atom.  {\bf (b)} Spontaneous emission of a three-leg giant atom with the same real-valued coupling parameter $c$ but different coupling distances $|x_3-x_2|=3|x_2-x_1|$ in the non-Markovian regime $\gamma \tau_{21}=0.25$ with decay rate $\gamma\equiv 2\Gamma c^2$ and propagating phase $\Omega\tau_{21}=\pi$.
   }\label{Fig-Pbreak}
\end{figure}

\section{Chirality and symmetries} 
For the real-valued coupling strength $c_m\in\mathbb{R}$, the system preserves the $\cal {T}$-symmetry, i.e., ${\cal {T}}\hat{H}_s{\cal{T}}^{-1}=\hat{H}_s^*=\hat{H}_s$ (see Appendix \ref{sec-symmetry}). The $\cal {T}$-symmetry can be broken by introducing  nonuniform phases $\theta_m$ to the coupling parameters $c_m=|c_m| e^{i\theta_m}$. Note that a uniform coupling phase $\theta_m=\theta_0$ can be removed by a displacement operation  ${\cal D}_l\hat{a} {{\cal D}_l}^{-1}=e^{ikl}\hat{a}_{k}$ (see Appendix \ref{sec-symmetry}). 
Here, we clarify that the chiral emission of a giant atom is not directly related to the  $\cal {T}$-symmetry breaking but to the $\cal{P}$-symmetry breaking, i.e. ${\cal{P}}\hat{H}_s {\cal{P}}^{-1}\neq\hat{H}_{s}$ . The chiral emission from $\cal{T}$-symmetry breaking is actually associated with the breaking of $\cal{P}$-symmetry. In fact, the $\cal{T}$-symmetry breaking is not a sufficient condition for chiral emission. For example, by designing a nonuniform configuration of coupling phases with mirror symmetry $\theta_m=\theta_{N-m+1}$, the $\cal{T}$-symmetry is broken while the $\cal{P}$-symmetry is preserved. In this scenario, we prove rigorously that the emission has no chirality $\mathcal{C}(t)=0$ at any time (see Appendix \ref{app-chiral}).

It is interesting to ask if the emission is chiral in the general case that the coupling parameters are real-valued but the coupling strengths or distances are nonuniform, where the $\cal{T}$-symmetry is still preserved while the $\cal{P}$-symmetry is broken.
In the Markovian regime, where the time delays among coupling points are neglected, the multi-leg giant atom is equivalent to a small atom with the atomic dynamics given by $\beta(t)=e^{-i(\Omega +\tilde{\Delta})t-\tilde{\gamma}t}\beta(0).$
Here, $\tilde{\Delta}$ is the Lamb shift and $\tilde{\gamma}$ is the effective decay rate depending on the details of coupling legs.  Combined with Eq.~(\ref{eq-varphi}), we find the phases of the emitted field in two directions are related by $\varphi(x_1,t)=e^{i\Omega |x_N-x_1|/v}\varphi^*(x_N,t)$. %%%Is this right?
In the non-Markovian regime, where the time delays among coupling points cannot be neglected, we prove rigorously that the chirality exists during finite time evolution $\mathcal{C}(t<\infty)\neq 0$, but vanishes in the infinitely long time limit $\mathcal{C}(\infty)= 0$ (see Appendix \ref{app-chiral}). In Fig.~\ref{Fig-Pbreak}, we show the time evolutions of the accumulated probabilities along two directions and the chirality for the two cases of a two-leg giant atom with different coupling strengths, and a three-leg giant atom with identical coupling strengths but different distances.
In a word, the $\cal{T}$-symmetry breaking is not a necessary condition for the chiral emission.

\begin{figure}
  \centering
 \centerline{\includegraphics[width=\linewidth]{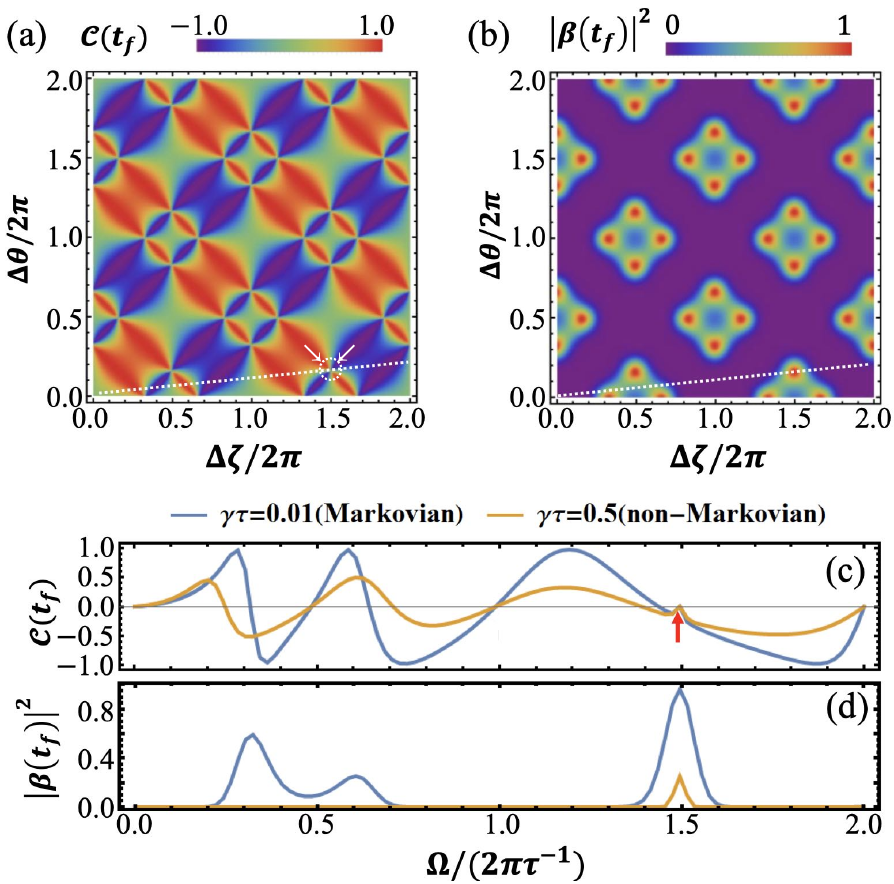}}
  \caption{{\bf  Chirality diagram. } 
  {\bf (a)} Chirality $\mathcal{C}(t_f)$ and {\bf (b)} residual atomic excitation $|\beta(t_f)|^2$ with $t_f/\tau=100\pi$ of the passive giant-atom architecture, shown in Fig.~\ref{Fig-system}(b), as a function of the propagating phase $\Delta\zeta=\Omega \tau$ with $\tau=d/v$ and the coupling phase $\Delta\theta=\Omega\Delta l/v$ with $N=3$ coupling points. 
 {\bf (c)} Chirality $\mathcal{C}(t_f)$  and {\bf (d)} residual atomic excitation $|\beta(t_f)|^2$ as a function of the atomic frequency $\Omega$ along the dashed white line shown in (a) and (b). We compare the results in the Markovian regime $\gamma\tau=0.01$ (blue) and that in the non-Markovian regime $\gamma\tau=0.5$ (yellow) with $\tau=d/v$. 
   }\label{Fig-chiral}
\end{figure}

\section{Passive coupling phases}
To have chirality in the infinitely long-time limit, we propose another scenario of breaking $\mathcal{P}$-symmetry while preserving $\mathcal{T}$-symmetry.
As sketched in Fig.~\ref{Fig-system} (b), we extend the coupling legs of the giant atom via meandering transmission lines to the 1D waveguide such that the propagating photon also accumulates phases through the legs. Different from the previous proposals with active inductive coupling~\cite{Wang2022qst}, the giant atom in our proposal is still capacitively coupled to the waveguide. Assuming the coupling leg has length $l_j$, the propagating photon accumulates a phase factor given by $e^{i\theta_m}$ with $\theta_m=\Omega l_m/v$.
By making the lengths of coupling legs nonuniform, the whole system breaks the $\mathcal{P}$-symmetry while the $\mathcal{T}$-symmetry is still preserved.
The effective Hamiltonian is equivalent to the model described in Eq.~(\ref{TotalH}) with the phases in the complex coupling parameter given by $\theta_m$. 
The emission of the giant atom can be tuned to be chiral depending on the choice of coupling phases. Compared to the previous proposal using sophisticated manipulation of active circuit elements \cite{joshi2023prx}, our proposal here is a passive superconducting circuit architecture without the need for a time-modulated JJ circuit that bears the advantage of resilience to noises.

\section{Chirality diagram}
We now investigate a designed architecture of $N$ uniformly distributed coupling points with a constant distance $d$. The accumulated propagating phases of the photon between the neighboring legs are $\Delta\zeta\equiv\Omega d/v=\Omega\tau$. Without loss of generality, we assume the coupling strength of each leg is a constant $|c_m|=c$ and the length difference between neighboring legs is also a constant $\Delta l$. As a result, the coupling phase varies in the arithmetic sequence $\theta_m =\theta_1+(m-1)\Delta \theta$ with $\Delta \theta\equiv\Omega\Delta l/v$.
In the Markovian regime, we neglect the traveling times of any two coupling points  
%%%\tilde{\beta}is not  defined here
while keep the accumulated phases \cite{Guo2017}. As a result, we have the dynamics of giant atom $\beta(t)=e^{-i(\Omega +\tilde{\Delta})t-\tilde{\gamma}t}$ with the Lamb shift $\tilde{\Delta}$ and the effective decay rate $\tilde{\gamma}$ given by(see Appendix \ref{App-effective})
\begin{eqnarray}\label{eq-lamb-effdecay}
 \left \{ \begin{array}{lll}
\tilde{\Delta}&=&-\frac{1}{4}\gamma\sum_{j=\pm}\frac{N\sin\phi_j-\sin (N\phi_j)}{1-\cos\phi_j},\\
\tilde{\gamma}&=&\frac{1}{4}\gamma\sum_{j=\pm}\frac{1-\cos(N\phi_j)}{1-\cos \phi_j},
 \end{array} \right.
\end{eqnarray}
where we have defined $\gamma\equiv2\Gamma c^2$ and $\phi_{\pm}\equiv\Delta\zeta\pm\Delta\theta$. 
%%%\gamma is not difiend here
%
%
The effective decay rate $\tilde{\gamma}$ takes the maximum $\tilde{\gamma}_{max}=N^2\gamma/2$ when the two conditions, i.e., $\phi_+=2k_+\pi$ with $k_+\in \mathbb{Z}$ and $\phi_-=2k_-\pi$ with $k_-\in \mathbb{Z}$, are both satisfied. The effective decay rate is taken to be zero $\tilde{\gamma}=0$ (the giant atom is decoupled to the waveguide) when the condition $\phi_{\pm}={2k_\pm\pi}/{N}$ for $k_\pm\in \mathbb{Z}$ and $k_\pm/N\notin \mathbb{Z}$ is satisfied.

By taking $\tilde{\gamma}\tau\ll1$ and $\tilde{\Delta}\tau\ll1$, we obtain the accumulated probabilities to the left and right directions 
\begin{eqnarray}
\mathcal{I}_{L/R}=\frac{\sin^2\frac{\phi_{+/-}}{2}\sin^2\frac{N\phi_{-/+}}{2}}{\sin^2\frac{\phi_{+/-}}{2}\sin^2\frac{N\phi_{-/+}}{2}+\sin^2\frac{\phi_{-/+}}{2}\sin^2\frac{N\phi_{+/-}}{2}},\label{eq-IL}\ \\\nn
\end{eqnarray}
and the chirality in the infinite long-time limit
\begin{eqnarray}\label{eq-chiralityNleg}
\mathcal{C}(\infty)=\frac{\sin^2\frac{\phi_+}{2}\sin^2\frac{N\phi_-}{2}-\sin^2\frac{\phi_-}{2}\sin^2\frac{N\phi_+}{2}}{\sin^2\frac{\phi_+}{2}\sin^2\frac{N\phi_-}{2}+\sin^2\frac{\phi_-}{2}\sin^2\frac{N\phi_+}{2}}.\ \ 
\end{eqnarray}
The full positive chirality $\mathcal{C}(\infty)=1$ is obtained, i.e., the emission to the right direction is forbidden ($\mathcal{I}_R=0$), when the phases $\Delta\theta$ and $\Delta\zeta$ satisfy the condition $\phi_+={2k_+\pi}/{N}$ with $k_+\in \mathbb{Z}$ and $k_+/N\notin \mathbb{Z}$. The full negative chirality $\mathcal{C}(\infty)=-1$ is obtained, i.e., the emission to the left direction is forbidden ($\mathcal{I}_L=0$), when the condition $\phi_-={2k_-\pi}/{N}$ with $k_-\in \mathbb{Z}$ and $k_-/N\notin \mathbb{Z}$ is satisfied.
The emissions to both directions are forbidden  ($\mathcal{I}_L=0$ and  $\mathcal{I}_R=0$) when the condition $\phi_{\pm}={2k_\pm\pi}/{N}$ with $k_\pm\in \mathbb{Z}$ and $k_\pm/N\notin \mathbb{Z}$ is satisfied. In this case, the giant atom is decoupled to the waveguide with zero effective decay rate ($\tilde{\gamma}=0$).

To verify Eq.~(\ref{eq-chiralityNleg}), in Fig.~\ref{Fig-chiral}(a), we plot the chirality $\mathcal{C}(t_f)$ with $t_f/\tau=100\pi$ as a function of coupling phase $\Delta\zeta $ and propagating phase $\Delta\theta$ directly from numerical simulations. In Fig.~\ref{Fig-chiral}(b), we plot the residual atomic excitation after a long time $|\beta(t_f)|^2$ as a function of two phases.
The chirality diagram shows $2\pi$-periodicity along both axes and is symmetric to the diagonal line. The numerical chirality $\mathcal{C}(t_f)$ is consistent with the analytical $\mathcal{C}(\infty)$ given by Eq.~(\ref{eq-chiralityNleg}) except the cusps as indicated by one representative dashed circle in the diagram, which correspond to the dark state points  ($\mathcal{I}_L=\mathcal{I}_R=0$). The chirality at the cusps is ill-defined as the limit of Eq.~(\ref{eq-chiralityNleg}) depends on the directions towards the cusps. In fact, the emission can be first tuned with opposite chiralities and then turned off gradually approaching the dark state points as indicated by two arrows in Fig.~\ref{Fig-chiral}(a). 
%
%In Fig.~\ref{Fig-chiral}(b), we plot the residual atomic excitation after a long time {\red $|\beta(t_f)|^2$} with $t_f/\tau=100\pi$ as a function of two phases. In the ideal case of an infinitely long time, only isolated points are located at the positions corresponding to the cusps in Fig.~\ref{Fig-chiral}(a). 

\section{Tunable chirality} 
According to the chirality diagram, we can tune the emission direction via the phase parameters $\Delta\zeta$ and $\Delta\theta$. In the proposal of active control strategies~\cite{zhang2021prl, Wang2022qst}, the phase parameter $\Delta\theta$ can be tuned independent of the phase parameter $\Delta\zeta$ via periodically modulating external JJ circuit loop. In our proposal of passive control strategy, the atomic transition frequency $\Omega$ can be used to tune the two phases $\Delta\zeta=\Omega d/v$ and $\Delta\theta=\Omega \Delta l/v$. Because the system parameters $d$ and $\Delta l$ cannot be tuned in situ, the ratio of the two phases $\lambda\equiv\Delta \theta/\Delta \zeta$ is fixed while the atomic frequency is tuned. Albeit this restriction, we show here that the chirality of emission can be adjusted on demand.

According to Eq.~(\ref{eq-IL}), the giant atom is in the dark state ($\mathcal{I}_L=\mathcal{I}_R=0$) when the parameters satisfy the condition $(1\pm \lambda)\Delta \zeta=2k_\pm\pi/N$ with $k_\pm\in \mathbb{Z}$ and $k_\pm/N\notin \mathbb{Z}$. Thus, the ratio $\lambda$ of two phases and the corresponding atomic frequency $\Omega=\Omega_c$ are given by
\begin{eqnarray}\label{eq-lambda}
\lambda=\frac{k_+-k_-}{k_++k_-}, \quad \Omega_c=\frac{(k_-+k_+)\pi v}{Nd}.
\end{eqnarray}
In Fig.~\ref{Fig-chiral}(a), we show the dashed lines that cross one dark state point for a three-leg giant atom corresponding to two integers $k_+=5$ and $k_-=4$ with the slope $\lambda=1/9$. In Fig.~\ref{Fig-chiral}(c), we then plot the chirality along the dashed line as a function of the atomic frequency $\Omega$. Indeed, the chirality can change from $\mathcal{C}=-1$ (complete right emission) to $\mathcal{C}=+1$ (complete left emission) via solely tuning atomic frequency. Again from Eq.~(\ref{eq-IL}), the complete left emission appears for $\Omega=k\Omega_c/k_+$ with $k\in \mathbb{Z}$, $k/k_+\notin\mathbb{Z}$ and $k/N\notin \mathbb{Z}$, while the complete right emission appears for $\Omega=k\Omega_c/k_-$, $k\in \mathbb{Z}$, $k/k_-\notin \mathbb{Z}$, $k/N\notin \mathbb{Z}$.
For the three-leg giant atom,  the complete left emission appears when $\Omega=\Omega_c/5, 2\Omega_c/5,4\Omega_c/5$, while the complete right emission is on the condition of $\Omega=\Omega_c/4, \Omega_c/2, 5\Omega_c/4$.
Note that the tip near point at $\Omega/(2\pi\tau^{-1})=1.5$ as indicated by the arrow reflects the fact that the chirality is ill-defined at the dark state point as discussed above.
%there are two different zero chirality points as indicated by the black and {\red red arrows in Fig.~\ref{Fig-chiral}(c)(upper)}. To illustrate this, we plot the residual population of the giant after a long time {\red $|\beta(t_f)|^2$ with $t_f=100\pi?$} {\red in Fig.~\ref{Fig-chiral}(c)(lower)}. For the zero chirality point at $\Delta\zeta/2\pi=1.0$, the atom completely decays into the waveguide. For the zero chirality point at $\Delta\zeta/2\pi=1.5$, the atom is in the dark state that does not decay to the waveguide. 
%
The non-Markovian effect, however, deteriorates the full chirality. In Fig.~\ref{Fig-chiral}(d), we plot the chirality as a function of atomic frequency for $\gamma\tau=0.5$, where the maximum value of chirality is lowered half compared to that for $\gamma\tau=0.01$ due to the leakage of the field from the legs before the constructive interference was established.
To avoid the non-Makrovian effect, we require the condition $\tilde{\gamma}_{max}\tau=N^2\gamma \tau/2\ll 1$ according to Eq.~(\ref{eq-lamb-effdecay}).%

\section{Nonreciprocity}
We have clarified above that the $\mathcal{T}$-symmetry breaking is neither a necessary nor a sufficient condition for the chirality of a giant atom. The chiral emission of a multi-leg giant atom is related to the $\mathcal{P}$-symmetry breaking. Now, we study the effects of symmetries on the reciprocity of the single-photon scattering for the giant atom by sending a single photon from the left (right) port of the waveguide and measuring the transmitted photon from the right (left) port of the waveguide. It is well-known that the scattering dynamics is \textit{nonreciprocal} if the $\mathcal{T}$-symmetry is broken for a linear system. However, we find here the $\mathcal{T}$-symmetry breaking for the giant atom via the coupling phases, cf. Fig.~\ref{Fig-system}(a), is \textit{artificial} because such a system can be mapped to the time-reversal system shown in Fig.~\ref{Fig-system}(b). Thus we would expect that the scattering is still reciprocal even if the coupling phases $\theta_m$ are nonuniform. The nonreciprocity can be only obtained via external dissipation of the giant atom that truly breaks the time-reversal symmetry. 

To further prove our point, we give the analytical expressions for the reflection coefficient $R$ and the transmission coefficient $T$ as follows (see details in Appendix \ref{App-scattering})
\begin{eqnarray}\label{}
 \left \{ \begin{array}{lll}
R&=&\Big|\frac{\Gamma\sum_{m,m'}c_mc^*_{m'}e^{i\omega_d(x_m+x_{m'})/v}}{\Delta+i\gamma_e+i\Gamma\sum_{m,m'}c_mc_{m'}^*e^{i\omega_d|x_{m}-x_{m'}|/v}}\Big|^2,\\
T&=&\Big|\frac{\Delta+i\gamma_e-2\Gamma\sum_{m<m'}c_mc_{m'}^*\sin\big(\omega_d\frac{|x_{m}-x_{m'}|}{v}\big)}{\Delta+i\gamma_e+i\Gamma\sum_{m,m'}c_mc_{m'}^*e^{i\omega_d|x_{m}-x_{m'}|/v}}\Big|^2,
 \end{array} \right.
\end{eqnarray}
where $\Delta=\omega_d-\Omega$ is the detuning between atomic and driving frequency. It can be analyzed that, under the mirroring operation $c_m\rightarrow c_{N-m+1}$ and $x_m\rightarrow x_N-x_m$, both reflection $R$ and transmission $T$ are invariant for zero external dissipation $\gamma_e=0$. However, for the finite external dissipation rate $\gamma_e>0$, the reflection $R$ is still invariant but the transmission $T$ is changed.
Considering the model of two coupling legs, with coupling coefficient $c_1=ce^{i\theta_1}$ and $c_2=ce^{i\theta_2}$. Assuming $\theta=\theta_2-\theta_1$ and $\omega_d|x_2-x_1|/v=\phi$, we have the reflection coefficient $R=\big|\frac{\Gamma c^2(1+2\cos\theta e^{i\phi}+e^{i2\phi})}{\Delta+i\gamma_e+2i\Gamma c^2(1+\cos\theta e^{i\phi})}\big|^2$ and transmission coefficient $T=\big|\frac{\Delta+i\gamma_e-2\Gamma c^2e^{-i\theta}\sin\phi}{\Delta+i\gamma_e+2i\Gamma c^2(1+\cos\theta e^{i\phi})}\big|^2$.
%
%\begin{eqnarray}\label{}
 %\left \{ \begin{array}{lll}
%R&=&\Big|\frac{\Gamma c^2(1+2\cos\theta e^{i\phi}+e^{i2\phi})}{\Delta+i\gamma_e+2i\Gamma c^2(1+\cos\theta e^{i\phi})}\Big|^2,\\
%T&=&\Big|\frac{\Delta+i\gamma_e-2\Gamma c^2e^{i\theta}\sin\phi}{\Delta+i\gamma_e+2i\Gamma c^2(1+\cos\theta e^{i\phi})}\Big|^2.
 %\end{array} \right.
%\end{eqnarray}
%
It is clear that, under the mirroring operation ($\theta\rightarrow -\theta$), the reflection is invariant while the transmission becomes $T\rightarrow\big|\frac{\Delta+i\gamma_e-2\Gamma c^2e^{i\theta}\sin\phi}{\Delta+i\gamma_e+2i\Gamma c^2(1+\cos\theta e^{i\phi})}\big|^2$ that is different from the previous one for finite dissipation rate $\gamma_e>0$. Especially, when the reflection is chosen to keep zero by taking $\phi=(2k+1)\pi-\theta$ with $k\in\mathbb{Z}$, the photon can completely transmit from the left side to the right side while the opposite transmission is fully forbidden when the detuning parameter satisfies $\Delta=\Gamma c^2\sin(2\theta)$ with $\theta=\arcsin\sqrt{\gamma_e/2\Gamma c^2}$. The opposite nonreciprocity can be achived by taking parameters $\phi=(2k+1)\pi+\theta$ with $k\in\mathbb{Z}$ and $\Delta=-\Gamma c^2\sin(2\theta)$ with $\theta=\arcsin\sqrt{\gamma_e/2\Gamma c^2}$.
%
%Such architecture can be used to design single-photon transistors or circulators that are crucial nonreciprocal signal routing and processing components involved in microwave read-out chains for a variety of applications~\cite{sliwa2015prx}.

\section{Summary}
In this work, we have clarified the roles of symmetries in the chirality and reciprocity of a giant atom. The $\cal{T}$-symmetry breaking is neither a sufficient condition nor a necessary condition for the chirality of a giant atom. We showed how to introduce and tune chirality for that giant atom by breaking $\mathcal{P}$-symmetry while preserving $\mathcal{T}$-symmetry. We also pointed out that the $\mathcal{T}$-symmetry breaking for a giant atom via coupling phases is an artificial aspect that cannot introduce nonreciprocity to the giant atom.
% {\red Our work suggests that chiral devices like single-photon routers can be realized via the coupling phases while nonreciprocal devices like single-photon transistors and circulators need the assistance of the external dissipation of giant atoms.}

\bigskip

\textbf{Acknowledgements}

This work was supported by the National Natural Science Foundation of China (Grant No. 12475025).

%\bibliography{/Users/lzguo/References/Refs-All.bib}
%\bibliography{Refs-All.bib}

%

\onecolumngrid

\newpage

\appendix

\section{Equations of motion and their solutions}\label{sec-model}
As described in the main text, the total Hamiltonian of a two-level giant atom coupled to the waveguides with  $N$ multiple coupling points is described as
\begin{eqnarray}
\hat{H}_s&=&\hbar\Omega\hat{\sigma}^+\hat{\sigma}^-
+\int_{-\infty}^{+\infty}\  dk\  \hbar\omega_k\
\hat{a}_k^\dag\hat{a}_{k}\\
&&+\sum_{m=1}^{N}\int_{-\infty}^{+\infty}
\Big(c_{m}e^{ikx_{m}}\hat{a}_k\hat{\sigma}^++{\rm h.c.}\Big)\sqrt{\omega_k}dk.\nonumber
\end{eqnarray}
For the spontaneous emission, there is only one excitation either in the atomic state or in the waveguide due to the Jaynes-Cummings interaction.  We study the single excitation subspace of the full system. The total system state can thus be described by
\begin{eqnarray}\label{TotalWave}
|\Psi(t)\rangle&=&\beta(t)|e,{\rm vac}\rangle+\int dk  \alpha_k(t) a^\dagger_k|g,{\rm vac}\rangle.
\end{eqnarray}
Here,  $|e\rangle$ and $|g\rangle$ represent the excited and ground state of the atom, respectively. $|\rm{vac}\rangle$ is the vacuum state in the waveguide, and
the integral describes the state of a single boson propagating in the waveguide. From the Schor\"{o}dinger equation $i\hbar \partial /\partial t|\Psi(t)\rangle=H|\Psi(t)\rangle$, we have the dynamics for the giant atom and the propagating modes in the waveguide
\begin{eqnarray}\label{EOMb}
\frac{d}{dt}\beta(t)&=&
-i\Omega\beta(t)-\frac{i}{\hbar}\sum_{m}c_{m}\int_{-\infty}^\infty
e^{ikx_{m}}\alpha_k(t) \sqrt{\omega_k}dk\label{EOMb-1}\\
\frac{d}{dt}\alpha_k(t)&=& -i\omega_k \alpha_k(t)-i\frac{\sqrt{\omega_k}}{\hbar}\sum_mc_m^*e^{-ikx_m}\beta(t).\label{EOMb-2}
\end{eqnarray}
Formally integrating the EOM (\ref{EOMb-2}) of the field operator, we have
\begin{eqnarray}\label{akt}
\alpha_k(t)=e^{-i\omega_k t}\left[\alpha_k(0)-i\frac{\sqrt{\omega_k}}{\hbar}\sum_mc_m^*e^{-ikx_m}\int_0^tdt'e^{i\omega_kt'}\beta(t')\right].
\end{eqnarray}
Inserting Eq.~(\ref{akt}) into Eq.~(\ref{EOMb}), we obtain
\begin{eqnarray}\label{bt0}
\frac{d}{dt}\beta(t)
&=&-i\Omega\beta(t)
-\frac{i}{\hbar}\sum_{m}c_{m}\int_{-\infty}^\infty \sqrt{\omega_k}
e^{ikx_{m}}e^{-i\omega_kt }\alpha_k(0) dk\nonumber\\
&&-\frac{1}{\hbar^2v}\sum_{m,m'}c_{m}c_{m'}^*\int_0^tdt'\beta(t')\int_{0}^\infty
e^{i\omega_k(|\tau_{mm'}|+t'-t)} \omega_kd\omega_k.
\end{eqnarray}
Here, we have used the linear dispersion relation of field modes $\omega_k=v|k|$ and $\tau_{mm'}\equiv (x_m-x_{m'})/v$ is the delay time between two coupling points at $x_m$ and $x_{m'}$.
In order to further simplify Eq.~(\ref{bt0}), we adopt the well-known Weisskopf-Wigner approximation\cite{scully1997quantum} for each coupling point. By approximating the integral over the frequency around  $\omega_k\approx\Omega$ and replacing $d\omega_k$ by $d(\omega_k-\Omega)$ in the last term of Eq.~(\ref{bt0}), we have

\begin{eqnarray}\label{Solution of EOM1}
\frac{d}{dt}\beta(t)
&\approx&-i\Omega\beta(t)
-i\sqrt{\frac{\Gamma v}{2\pi}}\sum_{m}c_{m}\int_{-\infty}^\infty
e^{ikx_{m}}e^{-i\omega_kt }\alpha_k(0) dk
-\Gamma\sum_{m,m'}c_mc_{m'}^*\int_0^t \delta(t-t'-|\tau_{mm'}|)\beta(t')dt'\nonumber\\
&=&-i\Omega\beta(t)
-i\sqrt{\frac{\Gamma v}{2\pi}}\sum_{m}c_{m}\int_{-\infty}^\infty
e^{ikx_{m}-i\omega_kt}\alpha_k(0) dk
-\Gamma
\sum_{m,m'}c_mc_{m'}^*\beta(t-|\tau_{mm'}|)\Theta(t-|\tau_{mm'}|).
\end{eqnarray}
Here, we have introduced the parameter $\Gamma\equiv\frac{2\pi\Omega}{\hbar^2
 v}$, and
$\Theta(x)$ is the Heaviside step function.
The EOM of $\beta(t)$ can be solved via Laplace transformation in the form of
\bea
\beta(t)=\chi(t)\beta(0)+\int_{-\infty}^\infty dk \xi_k(t)\alpha_k(0),
\eea
where the coefficients are given by
\begin{eqnarray}
 \chi(t)&=&\sum_{n}\frac{e^{s_n
t}}{1-\Gamma\sum_{m,m'}c_mc_{m'}^*|\tau_{mm'}|e^{-s_n|\tau_{mm'}|}},\label{PolesI}\\
\xi_k(t)&=&-i\sqrt{\frac{\Gamma v}{2\pi}}\sum_n\
 \frac{\sum_{m}c_m
 e^{ikx_m}e^{s_nt}}{(s_n+i\omega_k)[1-\Gamma\sum_{m,m'}c_mc_{m'}^*|\tau_{mm'}|e^{-s_n|\tau_{mm'}|}]}\nonumber\\
&&
-i\sqrt{\frac{\Gamma v}{2\pi}}\
 \frac{\sum_mc_m
 e^{ikx_m}e^{-i\omega_kt}}{i(\Omega-\omega_k)+\Gamma\sum_{m,m'}c_mc_{m'}^*e^{i\omega_k|\tau_{mm'}|}}\label{PolesII}
\end{eqnarray}
with $s_n$ in the upper equations the roots  of the following equation
\begin{eqnarray}\label{polesEq}
s+i\Omega+\Gamma\sum_{m,m'}c_mc_{m'}^*e^{-s|\tau_{mm'}|}=0.
\end{eqnarray}
Given the solution of $\beta(t)$, we obtain the total time-dependent field function in the waveguide $\varphi(x,t)$ at position $x$ and time $t$ in the transmission line from Eq.(\ref{akt})
\begin{eqnarray}\label{numberdensity}
\varphi(x,t)
&\equiv&\frac{1}{\sqrt{2\pi}}\int_{-\infty}^\infty dk
 e^{ikx}\ \alpha_k(t)\nonumber\\
&\approx&-i\sqrt{\frac{\Gamma }{v}} \sum_{m}c_{m}^*\beta(t-\frac{|x-x_m|}{v})\Theta(t-\frac{|x-x_m|}{v})+\frac{1}{\sqrt{2\pi}}\int_{-\infty}^\infty dk
 e^{i(kx-\omega_kt)}\ \alpha_k(0).
\end{eqnarray}
Here, we have used the Weisskopf-Wingner approximation again as we did in Eq.~(\ref{bt0}). The density of field in the transmission line is given by $p(x,t)=| \varphi(x,t)|^2$, which also describes the probability density to find a photon/phonon with all the possible wavenumber $k$.

For the spontaneous emission from the giant atom into the waveguide, i.e., $\alpha_k(0)=0, \beta(0)=1$, the EOM for the atomic probability amplitude $\beta(t)$ and the field function in the waveguide $\varphi(x,t)$ are simplified into
\begin{eqnarray}
\frac{d}{dt}\beta(t)
&=&-i\Omega\beta(t)
-\Gamma
\sum_{m,m'}c_mc_{m'}^*\beta(t-|\tau_{mm'}|)\Theta(t-|\tau_{mm'}|)\,,\nonumber\\
\beta(t)&=&\chi(t)=\sum_{n}\frac{e^{s_n
t}}{1-\Gamma\sum_{m,m'}c_mc_{m'}^*|\tau_{mm'}|e^{-s_n|\tau_{mm'}|}}\,,\nonumber\\
\varphi(x,t)
&=&-i\sqrt{\frac{\Gamma }{v}} \sum_{m}c_{m}^*\beta(t-\frac{|x-x_m|}{v})\Theta(t-\frac{|x-x_m|}{v}).
\end{eqnarray}

\section{Symmetry operations for a giant atom} \label{sec-symmetry}
We now discuss the operations of symmetries for a giant atom coupled to a linear waveguide. We first express the eigenstates of the free bosonic field modes in the linear waveguide by $|k\rangle=\int dx e^{ikx}|x\rangle$ and define the ladder operator of bosonic mode via $\hat{a}^\dagger_k |vac\rangle\equiv|k\rangle$. By operating the time-reversal operator ($\cal{T}$) on the eigenstate, i.e., ${\cal T}|k\rangle=\int dx e^{-ikx}|x\rangle=|-k\rangle=\hat{a}^\dagger_{-k}|vac\rangle$, we have the identity
\bea
{\cal{T}}\hat{a}^\dagger_k {\cal{T}}^{-1}{\cal{T}}|vac\rangle={\cal{T}}|k\rangle\ \ \  \Longrightarrow\ \ \ {\cal{T}}\hat{a}^\dagger_k {\cal{T}}^{-1}|vac\rangle=\hat{a}^\dagger_{-k}|vac\rangle\ .
\eea
As a result, we have
\bea
 {\cal{T}}\hat{a}^\dagger_k {\cal{T}}^{-1}=\hat{a}^{\dagger *}_k=\hat{a}^\dagger_{-k}.
\eea
For the mirroring or parity operator (${\cal P}$), we have ${\cal P}|k\rangle=\int dx e^{-ikx}|x\rangle=|-k\rangle=\hat{a}^\dagger_{-k}|vac\rangle$ and thus
\bea
 {\cal{P}}\hat{a}^\dagger_k {\cal{P}}^{-1}=\hat{a}^\dagger_{-k}.
\eea
For the translation (displacement) operator (${\cal D}_l$), we have ${{\cal D}_l}|k\rangle=\int dx e^{ik(x-l)}|x\rangle=e^{-ikl}|k\rangle=e^{-ikl}\hat{a}^\dagger_{k}|vac\rangle$ and thus
\bea\label{eq-sm-Dl}
 {\cal D}_l\hat{a}^\dagger_k {{\cal D}_l}^{-1}=e^{-ikl}\hat{a}^\dagger_{k}.
\eea
%Note that the above symmetry operations depend on the eigenstates in the waveguide, e.g., they are not valid for semi-infinite waveguide.

\section{Chirality of the spontaneous emission} \label{app-chiral}
We discuss the chirality of the spontaneous emission of the giant atom coupled to the waveguide with arbitrarily distributed coupling points and arbitrary coupling parameters.
Firstly, we consider the giant-atom system with ${\cal P}$-symmetry where the coupling legs are invariant by mirroring operation, i.e., $c_m=c_{N+1-m}$ and $x_m-x_1=x_N-x_{N+1-m}$ with $m=1,2,\cdots, N$.
According to Eq.~(\ref{EOMb-1}), the field function to the leftmost leg is given by
\begin{eqnarray}
\varphi(x_1,t)
&=&-i\sqrt{\frac{\Gamma }{v}} \sum_{m=1}^Nc_{m}^*\beta(t-\frac{|x_1-x_m|}{v})\Theta(t-\frac{|x_1-x_m|}{v}).
\end{eqnarray}
Similarly, the field function to the rightmost leg is 
\begin{eqnarray}
\varphi(x_N,t)
&=&-i\sqrt{\frac{\Gamma }{v}} \sum_{m=1}^Nc_{m}^*\beta(t-\frac{|x_N-x_m|}{v})\Theta(t-\frac{|x_N-x_m|}{v})\nonumber\\
&=&-i\sqrt{\frac{\Gamma }{v}} \sum_{m^\prime=N}^1c_{N+1-m^\prime}^*\beta(t-\frac{|x_N-x_{N+1-m^\prime}|}{v})\Theta(t-\frac{|x_N-x_{N+1-m^\prime}|}{v})\nonumber\\
&=&-i\sqrt{\frac{\Gamma }{v}} \sum_{m^\prime=1}^Nc_{m^\prime}^*\beta(t-\frac{|x_1-x_{m^\prime}|}{v})\Theta(t-\frac{|x_1-x_{m^\prime}|}{v})\nonumber\\
&=&\varphi(x_1,t).
\end{eqnarray}
The above derivation rigorously proves that for the ${\cal P}$-symmetric system, the field functions propagating to the left and right directions are equal, i.e., $\varphi(x_1,t)=\varphi(x_N,t)$, and thus the emission has no chirality at any time.
Especially, we can design a nonuniform but mirroring configuration of coupling phase $\theta_m=\theta_{N-m+1}$, where the time-reversal symmetry is broken while the mirroring symmetry is preserved. The fact of the absence of chirality in this case clarifies that the $\mathcal{T}$-symmetry breaking is not a sufficient condition for the chiral emission of a giant atom.

Secondly, we consider a more general condition for the giant atom without any symmetry. The probability of finding a photon emitted to the left (right) direction is the integral of the field density along the waveguide from the leftmost (rightmost) leg $x_1$ ($x_N$) to negative (positive) infinity, which is equal to time accumulation of the field density near the leftmost (rightmost ) leg $x_1$ ( $x_N$), i.e.,
\begin{eqnarray}
\mathcal{I}_L(t_f)&=&\int_{-\infty}^{x_1} |\varphi(x,t_f)|^2 dx=\int_{x_{fL}}^{x_1} |\varphi(x,t_f)|^2 dx=v\int_0^{t_f} |\varphi(x_1,t)|^2 dt \nonumber\\
&=&\Gamma\int_0^{t_f}|\sum_m c_m^*\beta(t-\frac{|x_1-x_m|}{v})\Theta(t-\frac{|x_1-x_m|}{v})|^2dt \nonumber\\
&=&\Gamma\sum_m |c_m|^2\int_0^{t_f-\frac{x_m-x_1}{v}}|\beta(t)|^2 dt\nonumber\\
&+&\Gamma\sum_{m> m^\prime}\left[c_m^*c_{m^\prime}\int_0^{t_f-\frac{x_m-x_1}{v}}\beta(t)\beta^*(t+\frac{x_m-x_{m^\prime}}{v})+\rm H.c.\right],\\
\mathcal{I}_R(t_f)&=&\int^{+\infty}_{x_N} |\varphi(x,t_f)|^2 dx=\int^{x_{fR}}_{x_N} |\varphi(x,t_f)|^2 dx=v\int_0^{t_f} |\varphi(x_N,t)|^2 dt \nonumber\\
&=&\Gamma\int_0^{t_f}|\sum_m c_m^*\beta(t-\frac{|x_N-x_m|}{v})\Theta(t-\frac{|x_N-x_m|}{v})|^2dt\nonumber\\
&=&\Gamma\sum_m |c_m|^2\int_0^{t_f-\frac{x_N-x_m}{v}}|\beta(t)|^2 dt\nonumber\\
&+&\Gamma\sum_{m> m^\prime}\left[c_m^*c_{m^\prime}\int_0^{t_f-\frac{x_N-x_m^\prime}{v}}\beta^*(t)\beta(t+\frac{x_m-x_{m^\prime}}{v}) dt+\rm H.c.\right].
\end{eqnarray} 
Here, $x_{fL}=x_1-vt$ ($x_{fR}=x_N+vt$) is the furthest position for the field propagating to the left (right) direction. 
The difference between the two probabilities is
\begin{eqnarray}
\mathcal{I}_L(t_f)-\mathcal{I}_R(t_f)
&=&\Gamma\sum_m |c_m|^2\int_{t_f-\frac{x_N-x_m}{v}}^{t_f-\frac{x_m-x_1}{v}}|\beta(t)|^2 dt\nonumber\\
&+&\Gamma\sum_{m> m^\prime}|c_mc_{m^\prime}|\left[e^{-i(\theta_m-\theta_{m^\prime})}\int_0^{t_f-\frac{x_m-x_1}{v}}\beta(t)\beta^*(t+\frac{x_m-x_{m^\prime}}{v}) dt+\rm H.c.\right]\nonumber\\
&+&\Gamma\sum_{m> m^\prime}|c_mc_{m^\prime}|\left[e^{i(\theta_m-\theta_{m^\prime})}\int_0^{t_f-\frac{x_N-x_m^\prime}{v}}\beta(t)\beta^*(t+\frac{x_m-x_{m^\prime}}{v}) dt+\rm H.c.\right]\nonumber\\
&=&\Gamma\sum_m |c_m|^2\int_{t_f-\frac{x_N-x_m}{v}}^{t_f-\frac{x_m-x_1}{v}}|\beta(t)|^2 dt\nonumber\\
&+&\Gamma\sum_{m> m^\prime}|c_mc_{m^\prime}|\left\{\left[e^{-i(\theta_m-\theta_{m^\prime})}-e^{i(\theta_m-\theta_{m^\prime})}\right]\int_0^{t_f-\frac{x_m-x_1}{v}}\beta(t)\beta^*(t+\frac{x_m-x_{m^\prime}}{v}) dt+\rm H.c.\right\}\nonumber\\
&+&\Gamma\sum_{m> m^\prime}|c_mc_{m^\prime}|\left[e^{i(\theta_m-\theta_{m^\prime})}\int_{t_f-\frac{x_N-x_m^\prime}{v}}^{t_f-\frac{x_m-x_1}{v}}\beta(t)\beta^*(t+\frac{x_m-x_{m^\prime}}{v}) dt+\rm H.c.\right].
\end{eqnarray}
In the general condition without $\mathcal{P}$-symmetry, the probability difference is not zero $\mathcal{I}_L(t_f)-\mathcal{I}_R(t_f)\neq 0$ during finite time evolution. 
In the long time limit, we assume the residual atomic excitation is zero $\beta(\infty)= 0$ except for the dark state points. Then, the probability difference for detecting the photon along two directions is simplified as
\begin{eqnarray}
\mathcal{I}_L(\infty)-\mathcal{I}_R(\infty)
&=&
\Gamma\sum_{m> m^\prime}|c_mc_{m^\prime}|\left\{\left[e^{-i(\theta_m-\theta_{m^\prime})}-e^{i(\theta_m-\theta_{m^\prime})}\right]\int_0^\infty\beta(t)\beta^*(t+\frac{x_m-x_{m^\prime}}{v}) dt+\rm H.c.\right\}\nonumber\\
&=&4\Gamma\sum_{m> m^\prime}|c_mc_{m^\prime}|\sin(\theta_m-\theta_{m^\prime})\int_0^\infty {\rm Im}\left[\beta(t)\beta^*(t+\frac{x_m-x_{m^\prime}}{v})\right] dt.
\end{eqnarray}
Interestingly, we have $\mathcal{I}_L(\infty)-\mathcal{I}_R(\infty)=0$ for the constant coupling phases, i.e., $\theta_m-\theta_{m^\prime}=0$ for arbitrary $m, m^\prime$. Note that a uniform coupling phase $\theta_m=\theta_0$ can be removed by a displacement operation  ${\cal D}_l\hat{a} {{\cal D}_l}^{-1}=e^{ikl}\hat{a}_{k}$, cf. Eq.~(\ref{eq-sm-Dl}). 
We thus conclude that the emission chirality exists for the finite time evolution but disappears in the long time limit for the real-valued coupling ($\mathcal{T}$-symmetry) strengths without $\mathcal{P}$-symmetry. 
%
%When the coupling phase are not mirror symmetric distributed, the stable (long time limit) chirality emerges.

%\section{Proof of the equal density flux for mirror symmetric system} \label{AppC}

\section{Derivation for the effective model} \label{App-effective}
In the Markovian regime, the giant atom with multiple legs can be simplified into an effective model, in which the $N$ legs
can be treated together as a single coupling point with the effective coupling strength that the system parameters can tune. In this section, we will derive the effective model in detail from the EOM of the giant atom.

As shown in Fig.~1 in the main text, we consider a designed architecture of $N$ uniformly distributed coupling points with a constant distance $d$, and time delay $\tau$ between neighboring points.
Therefore, all the possible delay time can be represented as $|\tau_{mm'}|=n\tau$ with $n=0,1,\cdots,N-1$.  The combination number of the time delays is $N$ for $n=0$ and $2(N-n)$ for $n\neq 0$.
Without loss of generality, we assume the coupling strength of each leg is a constant $|c_m|=c$ and the coupling phase varies in the arithmetic sequence $\theta_m =\theta_1+(m-1)\Delta \theta$, with $m=1,2,\cdots,N$.
Therefore, we have the EOM of $\beta(t)$ from Eq.~(\ref{EOMb-1}), 
\begin{eqnarray}
\frac{d}{dt}\beta(t)
&=&-i\Omega\beta(t)
-\Gamma
\sum_{m,m'}c_mc_{m'}^*\beta(t-|\tau_{mm'}|)\Theta(t-|\tau_{mm'}|)\,,\nonumber\\
&=&-(i\Omega+\frac{1}{2}N\gamma)\beta(t)-\gamma\sum_{n=1}^{N-1}(N-n)\cos(n\Delta\theta)\beta(t-n\tau)\Theta(t-n\tau)\,,
\end{eqnarray}
with parameter $\gamma=2\Gamma c^2$. Note that above EOM is written in the lab frame. In the RWA, the dynamics of $\beta(t)$
 can be separated into a fast oscillation with frequency $\Omega$ and a slow time evolution of amplitude $\tilde{\beta}(t)$, i.e., $\beta(t)=\tilde{\beta}(t)e^{-i\Omega t }$.
As a result, the EOM of $\tilde{\beta}(t)$ has the form of
\begin{eqnarray}
\frac{d}{dt}\tilde{\beta}(t)
&=&-\frac{1}{2}N\gamma\tilde{\beta}(t)
-\frac{1}{2}\gamma\sum_{n=1}^{N-1}(N-n)[e^{in(\Delta\theta+\Omega\tau)}+e^{in(-\Delta\theta+\Omega\tau)}]\tilde{\beta}(t-n\tau)\Theta(t-n\tau)\,.
\end{eqnarray}
By neglecting the time delays $n\tau$ in the slow amplitude $\tilde{\beta}(t-n\tau)$ in the Markovian regime, we have
\begin{eqnarray}\label{EOMtilbeta}
\frac{d}{dt}\tilde{\beta}(t)
&\approx&-\frac{1}{2}N\gamma\tilde{\beta}(t)
-\frac{1}{2}\gamma\sum_{n=1}^{N-1}(N-n)[e^{in(\Delta\theta+\Omega\tau)}+e^{in(-\Delta\theta+\Omega\tau)}]\tilde{\beta}(t)\nonumber\\
&=&-\sum_{j=\pm}\frac{1}{2}\gamma\left[\frac{1-\cos
(N\phi_j)}{2(1-\cos \phi_j)}+i\frac{N\sin\phi_j-\sin (N\phi_j)}{2(1-\cos
\phi_j)}\right]\tilde{\beta}(t)\,
\end{eqnarray}
with paramters $\phi_{\pm}=\Omega\tau\pm\Delta\theta$. Then, we extract the Lamb shift  $\tilde{\Delta}$ and the effective decay rate $\tilde{\gamma}$ as
\begin{eqnarray}
\tilde{\Delta}&=&-\frac{1}{4}\gamma\sum_{j=\pm}\frac{N\sin\phi_j-\sin (N\phi_j)}{1-\cos
\phi_j},\nonumber\\
\tilde{\gamma}&=&\frac{1}{4}\gamma\sum_{j=\pm}\frac{1-\cos
(N\phi_j)}{1-\cos \phi_j}\,.
\end{eqnarray}
Using the relationship $\tilde{\beta}(t)=e^{i\Omega t }\beta(t)$, we have the EOM of the lab-frame amplitude $\beta(t)$ as follows
\begin{eqnarray}\label{AppEOM2IDT-1}
\frac{d}{dt}\beta(t)&=&-i(\Omega+\tilde{\Delta})\beta(t)
-\tilde{\gamma}\beta(t).
\end{eqnarray}
The above EOM (\ref{AppEOM2IDT-1}) describes the simplified effective small-atom model with one single coupling point from  the
giant-atom model with $N$ coupling points in the Markovian regime.

\section{Nonreciprocal scattering for giant atom}\label{App-scattering}
In this section, we will investigate the reciprocity of the single-photon scattering for the giant atom by sending a single photon from one port of the waveguide and measuring the transmitted/reflected photons.
In the calculation of the scattering coefficients, we 
follow the linear response method developed in Ref. \cite{Guo2024njp} by introducing a weak driving term in the Hamiltonian as a probe field. We solve analytically the response of the giant-atom system and consequently derive the scattering coefficients.

%we introduce an external dissipation to the two-level giant atom, with rate $\gamma_e$,

%we follow the method in Ref. \cite{Guo2024njp}, and consider the model of 
%scattering dynamics of giant atom and bosonic field in the transmission line for the 

%To calculate of the scattering coefficients,

%consider the coupling phase of each leg, and review the dynamics of the scattering dynamics of giant atom and bosonic field in the transmission line, following the method in Ref. \cite{Guo2024njp}.

\subsection{Hamiltonian and Linear response}

\begin{figure}
  \centering
 \centerline{\includegraphics[width=0.7\linewidth]{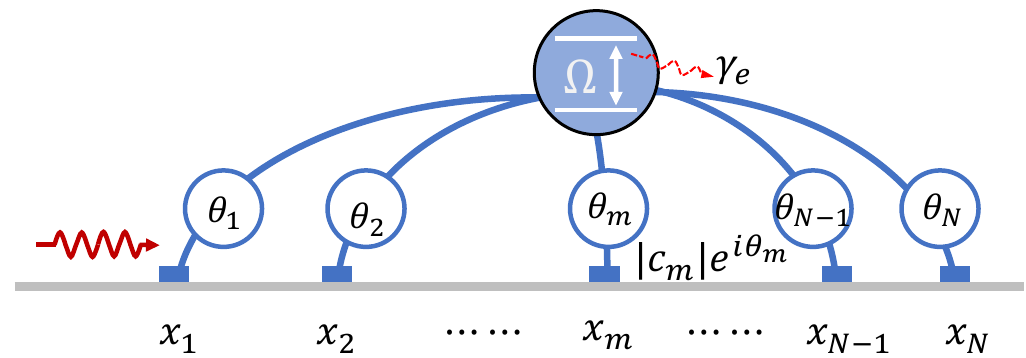}}
  \caption{{\bf Scattering for the giant atom.}
   A two-level atom (light blue) with transition frequency $\Omega$ and dissipation rate $\gamma_e$ coupled to a transmission line (gray) via multiple points $x_m$ with complex coupling strength $c_m=|c_m|e^{i\theta_m}$. An incident probe wave is sent from
the left to the right in the transmission line.   
   }\label{Figapp-system}
\end{figure}

As shown in Fig.~\ref{Figapp-system}, we consider a giant atom coupled to the transmission line via $N$ coupling points with an additional decay channel. We describe the total system with a non-Hermitian Hamiltonian given by
\cite{Guo2024njp, Chen2022cp}
\begin{eqnarray}
{H}_s&=&\hbar(\Omega-i\gamma_e)\hat{b}^\dagger\hat{b}
+\int_{-\infty}^{+\infty}\  dk\  \hbar\omega_k\
\hat{a}_k^\dag\hat{a}_{k}
+\sum_{m=1}^{N}\int_{-\infty}^{+\infty}
\Big(c_{m}e^{ikx_{m}}\hat{a}_k\hat{b}^\dagger+{\rm h.c.}\Big)\sqrt{\omega_k}dk.
\end{eqnarray}
The first non-Hermitian term on the right-hand side (RHS) describes the giant atom with transition frequency $\Omega$ and external dissipation rate $\gamma_e$. In the second term on the RHS, the field operators $\hat{a}_k$ represent the bosonic mode with wavenumber $k$ in the 1D waveguide satisfying $[\hat{a}_k,\hat{a}^\dagger_{k'}]=\delta(k-k')$. The waveguide has a linear dispersion relation $\omega_k=|k|v$ with the wavenumber $k$ and the propagating velocity $v$. In the third term on the RHS, the complex parameter $c_m=|c_m|e^{i\theta_m}$ represents the coupling strength at coupling position $x_m$.

We assume an incident probe plane wave with frequency $\omega_d$ and amplitude $f$ is sent from the
left to the right in the transmission line.
By setting the moment when the plane wave passes the leftmost coupling point as the initial time, the total Hamiltonian is $H=H_s+H_D$ with $H_D$ the incident driving probe Hamiltonian given by 
\begin{eqnarray}\label{HD}
H_D&=& \frac{i}{2}\hbar\Omega_f\hat{b}^\dagger+{\rm H.c.}=
\frac{i}{2}\hbar f \sum_m c_m e^{-i\omega_d(t-\tau_m)}\Theta(t-\tau_m)\hat{b}^\dagger+{\rm H.c.}
\end{eqnarray}
Here, the parameter $\Omega_f \equiv f\sum_{m}c_me^{-i\omega_d(t- \tau_m)}\Theta(t-\tau_m)$ describes the collective driving effect of all the coupling points, where $\tau_m\equiv
(x_m-x_1)/v$ is the traveling time of the drive and  $\Theta(x)$ is the Heaviside step function.

In order to solve the scattering problem in the presence of the probe drive and external dissipation, we use the Heisenberg
EOMs for the atomic operator $\hat{b}(t)$ and the field operator $\hat{a}_k(t)$ as follows
\begin{eqnarray}\label{EOMHb}
\frac{d}{dt}\hat{b}(t)&=&\frac{1}{i\hbar}[\hat{b}(t),H]
=-i(\Omega-i\gamma_e)\hat{b}(t)+\frac{\Omega_f}{2}-\frac{i}{\hbar}\sum_{m}c_{m}\int_{-\infty}^\infty
e^{ikx_{m}}\hat{a}_k(t) \sqrt{\omega_k}dk\label{EOMHb-1},\\
\frac{d}{dt}\hat{a}_k(t)&=& \frac{1}{i\hbar}[\hat{a}_k(t),H]=-i\omega_k \hat{a}_k(t)-i\frac{\sqrt{\omega_k}}{\hbar}\sum_mc_m^*e^{-ikx_m}\hat{b}(t).\label{EOMHb-2}
\end{eqnarray}
Without probe drive or dissipation ($\gamma_e=0$, $\Omega_f=0$), Eqs.~(\ref{EOMHb-1}) and (\ref{EOMHb-2}) have the exact same form as Eqs. (\ref{EOMb-1}) and (\ref{EOMb-2})
by replacing $\beta(t)$ and $\alpha_k(t)$ with operators $\hat{b}(t)$ and $\hat{a}_k(t)$, respectively. 
As already proved in Ref.~\cite{guo2020prr,Guo2024njp}, in the linear problem, the complex harmonic amplitude $\langle \hat{b}(t)\rangle$ follows exactly the same EOM as
that of the probability amplitude $\beta(t)$ for the two-level atom in the single-excitation approximation.
Therefore, we can study the single-photon scattering for the two-level giant atom based on the solution of the atomic operator $\hat{b}(t)$ and field operator $\hat{a}_k(t)$.
Since the EOM for $\hat{b}(t)$ is linear, we assume the solution has the following form
\begin{eqnarray}\label{Hbt}
 \hat{b}(t)&\equiv& \eta(t)+\chi(t)\hat{b}(0)+\int_{-\infty}^\infty dk
\xi_k(t)\hat{a}_k(0)
\end{eqnarray}
with initial conditions $\eta(0)=\xi_k(0)=0, \chi(0)=1$. 
Following the derivation similar to that in section \ref{sec-model}, we obtain the analytical solutions of $\eta(t), \chi(t)$ and $\xi_k(t)$ as follows
\begin{eqnarray}\label{etat}
\eta(t)&=&\frac{f\sum_mc_me^{-i\omega_d
(t-\tau_m)}}{2(-i\Delta+\gamma_e+\Gamma\sum_{m,m'}c_mc_{m'}^*e^{i\omega_d|\tau_{mm'}|})}\nonumber\\
&&+\sum_{n}\frac{f\sum_mc_me^{s_n(t-
\tau_m)}}{2(s_n+i\omega_d)(1-\Gamma\sum_{m,m'}c_mc_{m'}^*|\tau_{mm'}|e^{-s_n|\tau_{mm'}|})},\\
\chi(t)&=&\sum_{n}\frac{e^{s_n
t}}{1-\Gamma\sum_{m,m'}c_mc_{m'}^*|\tau_{mm'}|e^{-s_n|\tau_{mm'}|}},\\
\xi_k(t)&=&-i\sqrt{\frac{\Gamma v}{2\pi}}\sum_n\
 \frac{\sum_{m}c_m
 e^{ikx_m}e^{s_nt}}{(s_n+i\omega_k)[1-\Gamma\sum_{m,m'}c_mc_{m'}^*|\tau_{mm'}|e^{-s_n|\tau_{mm'}|}]}\nonumber\\
&&
-i\sqrt{\frac{\Gamma v}{2\pi}}\
 \frac{\sum_mc_m
 e^{ikx_m}e^{-i\omega_kt}}{i(\Omega-\omega_k)+\gamma_e+\Gamma\sum_{m,m'}c_mc_{m'}^*e^{i\omega_k|\tau_{mm'}|}}.
\end{eqnarray}
Here, $\eta(t)$ represents the linear response of the giant atom to the probe drive. 
The parameter $\Delta\equiv\omega_d-\Omega$ is the atom-drive frequency detuning, and $s_n$ represents the roots of the following equation
\begin{eqnarray}
s+i\Omega+\gamma_e+\Gamma\sum_{m,m'}c_mc_{m'}^*e^{-s|\tau_{mm'}|}=0.
\end{eqnarray}
Given the solution of atomic operator $\hat{b}(t)$,  the field operator can be obtained by integrating Eq.~(\ref{EOMHb-2}), i.e.,
\begin{eqnarray}\label{Hakt}
\hat{a}_k(t)=e^{-i\omega_k t}\left[\hat{a}_k(0)-i\frac{\sqrt{\omega_k}}{\hbar}\sum_mc^*_me^{-ikx_m}\int_0^tdt'e^{i\omega_kt'}\hat{b}(t')\right].
\end{eqnarray}

\subsection{Scattering dynamics of bosonic field}
We now discuss the scattering dynamics of bosonic field in the transmission line. To calculate the scattering coefficients, we define the following field operator
\begin{eqnarray}
\hat{\Psi}(x,t)
&\equiv&\int_{-\infty}^\infty dk\sqrt{\omega_k}
 e^{ikx}\ \hat{a}_k(t)\nonumber\\
 &=&-i\hbar \Gamma\sum_{m'}c_{m'}^*\hat{b}(t-|x-x_{m'}|/v)\Theta(t-|x-x_{m'}|/v)
+\int_{-\infty}^\infty
 dk\sqrt{\omega_k}
 e^{ikx_m}\ \hat{a}_k(0)e^{-i\omega_kt}.
\end{eqnarray}
Note that the above field operator $\hat{\Psi}(x,t)$, with a factor $\sqrt{\omega_k}$ in the integrand, is different from the previous field operator $\hat{\varphi}(x,t)$ defined in Eq.~(\ref{numberdensity}). The quantity of $|\langle\hat{\Psi}(x,t)\rangle|^2$ measures the intensity of field energy (scaled by $\hbar$) while the quantity of $|\langle\hat{\varphi}(x,t)\rangle|^2$ measures the density of bosonic mode number (or the probability for single photon/phonon dynamics) in the transmission line.
Assuming the initial state of giant atom and the bosonic field are both in the ground states, we then have $\langle \hat{b}(t)\rangle=\eta(t)$ according to Eq.~(\ref{Hbt}).

The reflection coefficient can be calculated from the
bosonic field energy intensity near the left side of the first coupling point $\langle\hat{\Psi}(x_1,t)\rangle$.
According to the driving Hamiltonian Eq.~(\ref{HD}),
the corresponding probe driving amplitude is $\frac{1}{2}i\hbar
fe^{-i\omega_dt}$.
Therefore, the dynamic of reflection coefficient is given by
\begin{eqnarray}\label{Rt}
R(t)\equiv\frac{|\langle\Psi(x_1,t)\rangle|^2}{|\frac{1}{2}\hbar fe^{-i\omega_dt}|^2}
&=&\frac{4\Gamma^2}{|f|^2}\Big|\sum_{m}c_{m}^*\eta(t-\tau_{m})\Big|^2.
\end{eqnarray}
According to Eq.~(\ref{etat}), the corresponding long-time limit of the reflection coefficient is given by
\begin{eqnarray}\label{Rlim}
R(\infty)&=&\Big|\frac{\Gamma\sum_{m,m'}c_mc_{m'}^*e^{i\omega_d(\tau_m+\tau_{m'})}}{\Delta+i\gamma_e+i\Gamma\sum_{m,m'}c_mc_{m'}^*e^{i\omega_d|\tau_{mm'}|}}\Big|^2\nonumber\\
&=&\frac{\Gamma^2|\sum_{m}c_me^{i\omega_d\tau_m}|^2|\sum_{m}c_m^*e^{i\omega_d\tau_m}|^2}{|\Delta+i\gamma_e+i\Gamma\sum_{m,m'}c_mc_{m'}^*e^{i\omega_d|\tau_{mm'}|}|^2}.
\end{eqnarray}
In contrast, the field near the right side of the $N$-th coupling point is the superposition of the
probe field and transmitted field.
Since there are $N$ coupling points in total,
the dynamics of transmission
coefficient is then given by
\begin{eqnarray}\label{Tt}
T(t)&\equiv&\frac{|\frac{1}{2}i\hbar fe^{-i\omega_d(t- \tau_N)}+\langle\hat{\Psi}(x_N,t)\rangle|^2}{|\frac{1}{2}\hbar fe^{-i\omega_dt}|^2}\nonumber\\
&=&\frac{|fe^{-i\omega_d(t- \tau_N)}-2\Gamma
\sum_{m'}c_{m'}^*\eta(t-\tau_N+\tau_{m'})|^2}{|fe^{-i\omega_dt}|^2}
\end{eqnarray}
with the corresponding long-time limit
\begin{eqnarray}\label{eq-sm-T}
T(\infty)
&=&\Big|1-\frac{\Gamma|\sum_{m}c_me^{i\omega_d\tau_m}|^2}{-i\Delta+\gamma_e+\Gamma\sum_{m,m'}c_mc_{m'}^*e^{i\omega_d|\tau_{mm'}|}}
\Big|^2\nonumber\\
&=&\Big|\frac{\Delta+i\gamma_e-2\Gamma\sum_{m<m'}c_mc^*_{m'}\sin(\omega_d|\tau_{mm'}|)}{\Delta+i\gamma_e+i\Gamma\sum_{m,m'}c_mc_{m'}^*e^{i\omega_d|\tau_{mm'}|}}
\Big|^2.
\end{eqnarray}
In the case that there is no dissipation in the giant atom ($\gamma_e=0$), one can prove that the long-time limit of the reflection
coefficient and transmission coefficient have the relationship of $R(\infty)+T(\infty)=1$. In fact, from Eq.~(\ref{eq-sm-T}) we have
\begin{eqnarray}\label{}
T(\infty)
&=&1-\frac{2\Gamma^2|\sum_{m}c_me^{i\omega_d\tau_m}|^2}{|-i\Delta+\Gamma\sum_{m,m'}c_mc_{m'}e^{i\omega_d|\tau_{mm'}|}|^2} \mathrm{Re}[\sum_{m,m'}c_mc_{m'}e^{i\omega_d|\tau_{mm'}|}]\nonumber\\
&&+\frac{\Gamma^2|\sum_{m}c_me^{i\omega_d\tau_m}|^4}{|-i\Delta+\Gamma\sum_{m,m'}c_mc_{m'}^*e^{i\omega_d|\tau_{mm'}|}|^2}\nonumber\\
&=&1-\frac{\Gamma^2|\sum_{m}c_me^{i\omega_d\tau_m}|^2|\sum_{m}c_m^*e^{i\omega_d\tau_m}|^2}{|-i\Delta+\Gamma\sum_{m,m'}c_mc_{m'}^*e^{i\omega_d|\tau_{mm'}|}|^2}=1-R(\infty),
\end{eqnarray}
where we have used the following relationship
 \begin{eqnarray}\label{}
{\rm Re}[\sum_{m,m'}c_mc_{m'}^*e^{i\omega_d|\tau_{mm'}|}]=|\sum_{m}c_me^{i\omega_d\tau_m}|^2+|\sum_{m}c_m^*e^{i\omega_d\tau_m}|^2.
\end{eqnarray} 

\subsection{Nonreciprocal scattering for two-legs model}
We now discuss the nonreciprocal scattering for the giant atom coupled to the transmission line via two coupling legs, with coupling parameter $c_1=ce^{i\theta_1}$ and $c_2=ce^{i\theta_2}$.
Assuming $\theta=\theta_2-\theta_1$, $\tau_2-\tau_1=\tau$, $\Omega_d\tau=\phi$, we have the reflection coefficient and transmission coefficient for the left-incident photon as follows
\begin{eqnarray}\label{RTlim2}
R_L&=&\Big|\frac{\Gamma c^2(1+2\cos\theta e^{i\phi}+e^{i2\phi})}{\Delta+i\gamma_e+2i\Gamma c^2(1+\cos\theta e^{i\phi})}\Big|^2,\\
T_{L\to R}
&=&\Big|\frac{\Delta+i\gamma_e-2\Gamma c^2e^{-i\theta}\sin\phi}{\Delta+i\gamma_e+2i\Gamma c^2(1+\cos\theta e^{i\phi})}\Big|^2.
\end{eqnarray}
%The result is consistent with that in Ref[Communications physics:Nonreciprocal and chiral single-photon scattering
%for giant atoms Eq.(6a)and (6b)](a little different,left incident result is equal to  that from right)
For the right-incident photon, the scattering coefficients can be simply calculated by the mirroring operation ($\theta\to-\theta$), 
\begin{eqnarray}\label{RTlim3}
R_R&=&\Big|\frac{\Gamma c^2(1+2\cos\theta e^{i\phi}+e^{i2\phi})}{\Delta+i\gamma_e+2i\Gamma c^2(1+\cos\theta e^{i\phi})}\Big|^2.\\
T_{R\to L}
&=&\Big|\frac{\Delta+i\gamma_e-2\Gamma c^2e^{i\theta}\sin\phi}{\Delta+i\gamma_e+2i\Gamma c^2(1+\cos\theta e^{i\phi})}\Big|^2.
\end{eqnarray}
Comparing the above scattering coefficients for the left and right incident photon, we find the reflection coefficients are completely identical, while the transmission coefficients are invariant only for the zero external dissipation $\gamma_e=0$.

We call the scattering is \textit{nonreciprocal} if the left-incident and right-incident transmission coefficient are not identical, i.e., $T_{L\to R}\neq T_{R\to L}$. We define the \textit{nonreciprocity} as
\begin{eqnarray}\label{eq-nonreciprocity}
\mathcal{NR}\equiv\frac{T_{L\to R}-T_{R\to L}}{T_{L\to R}+T_{R\to L}}.
\end{eqnarray}
The full positive nonreciprocity $\mathcal{NR}=+1$ (the right-incident transmission is  forbidden $T_{R\to L}=0$) is achieved when the detuning and external dissipation satisfy $\Delta=2\Gamma c^2 \cos\theta\sin\phi,\, \gamma_e=2\Gamma c^2\sin\theta\sin\phi$. 
The opposite nonreciprocity $\mathcal{NR}=-1$  (the left-incident transmission is  forbidden $T_{L\to R}=0$) is obtained when the detuning and external dissipation satisfy $\Delta=2\Gamma c^2 \cos\theta\sin\phi,\, \gamma_e=-2\Gamma c^2\sin\theta\sin\phi$.

Especially, we can set the reflection coefficient to be zero via the conditions $\phi=(2k+1)\pi\pm \theta$ with $k\in \mathbb{Z}$. 
With this restraint and for the finite dissipation rate $\gamma_e>0$, we obtain the full positive nonreciprocity by taking $\phi=(2k+1)\pi-\theta$, with the detuning parameter  $\Delta=\Gamma c^2 \sin(2\theta)$ and dissipation rate $ \gamma_e=2\Gamma c^2\sin^2\theta$.
In contrast, the full negative nonreciprocity is obtained by taking $\phi=(2k+1)\pi+\theta$ together with detuning $\Delta=-\Gamma c^2 \sin(2\theta)$ and dissipation rate $ \gamma_e=2\Gamma c^2\sin^2\theta$.

%\section*{References}

%\bibliography{/Users/lzguo/References/Refs-All.bib}
%\bibliographystyle{iopart-num}

%\bibliography{NcFT-Refs}

\end{document}